# CONTROL OF MULTISCALE SYSTEMS WITH CONSTRAINTS.
## 2. FRACTAL NUCLEAR ISOMERS AND CLUSTERS

S. Adamenko [1], V. Bolotov [2], V. Novikov [3]

[1] Electrodynamics Laboratory *Proton-21*
[2] V. Karazin National University of Kharkov
[3] The Institute of Electrophysics and Radiation Technologies of the National Academy of Sciences of Ukraine

We consider the influence of the Fermi statistics of nucleons on the binding energy of a new type of nuclear structures such as fractal nuclear clusters (fractal isomers of nuclei). It is shown that the fractal nuclear isomers possess a wide spectrum of binding energies that exceed, in many cases, the values known at the present time. The transition of the nuclear matter in the form of ordinary nuclei (drops of the nuclear fluid) in the state with the fractal structure or in the form of bubble nuclei opens new sources of energy and has huge perspectives. This transition is based on a new state of matter – collective coherently correlated state. It manifests itself, first of all, in the property of nonlocality of nuclear multiparticle processes. We develop a phenomenological theory of the binding energy of nuclear fractal structures and modify the Bethe - Weizsäcker formula for nuclear clusters with the mass number $A$, charge $Z$, and fractal dimension $D_f$. The consideration of fractal nuclear isomers allows one to interpret the experimental results on a new level of the comprehension of processes of the nuclear dynamics. The possibility to determine the fractal dimension of nuclear systems with the help of the method of nuclear dipole resonance for fractal isomers is discussed. The basic relations for fractal electroneutral structures such as the electron - nucleus plasma of fractal isomers are presented.

**Introduction.**

During many decades, the development of the fundamental and applied nuclear physics is referred to the priority trends of science and technology in many countries. The special interest in the development of nuclear physics is mainly related to the hope for that the nuclear power industry could become the most powerful source of energy.

All primary sources of energy in the Nature have the single base, namely the processes with a change of the binding energy of systems. This concerns the most spread sources of energy based on the transformation of the binding energy on the atomic and molecular levels, e.g., at the combustion of organic fuel, and the nuclear processes generated by a change of the binding energy of nucleons in the nuclei of atoms [1] at the running of nuclear reactions.

The most powerful sources of energy are those which use the binding energy of many-nucleon nuclear systems, because the density of this energy has a value (by modern representations) of the order of several MeV per nucleon, as distinct from the chemical energy, whose value is several eV per atom or molecule.

The significance of the quantities directly related to the binding energy (mass defect, packing coefficient [2]) for the comprehension of the nature of nuclear phenomena and processes becomes clear very rapidly. When nuclear physics originated, the structure of nuclei and the interactions between nucleons composing a nucleus were known only in the very general features. At that time, some attempts to clarify the structure of a nucleus were based on the analysis of available data on the masses of nuclei (and, hence, their mass defects).

In [3], Weizsäcker obtained a rather awkward phenomenological formula for the masses of nuclei on the basis of experimental data on mass defects and the binding energy of nuclei in the Thomas – Fermi approximation of self-consistent field with regard for the finite size of a nucleus (finite value of the surface energy). The formula includes the sum of contributions of the bulk energy, surface energy, and Coulomb energy and well represents the general dependence of the binding energy of nuclei on the parameters of a nucleus (the number of protons $Z$ and the number of neutrons $N = A - Z$ in a nucleus). In work [4], Bethe modified the formula for the binding energy (in MeV) to the commonly accepted form, where the meaning of terms is quite transparent (especially from the viewpoint of collective representations):

$$B(A,Z) = \left(c_0 - c_3\left(1 - \frac{2Z}{A}\right)^2\right)A - c_1 A^{2/3} - c_2 \frac{Z^2}{A^{1/3}} + \frac{c_p}{A^{1/2}}\begin{cases}1, & Z=2l, N=2k \\ 0, & A=2k+1 \\ -1, & Z=2l+1, N=2k+1\end{cases};$$

$$c_0 = 15.75,\ c_1 = 17.8,\ c_2 = 0.71,\ c_3 = 23.7,\ c_p \approx 12.0.\qquad(1)$$

The first term gives the bulk contribution of the strong interaction jointly with the so-called symmetry energy related to the Pauli principle. The second term is the contribution of the surface of a nucleus to the binding energy. The third term corresponds to the Coulomb energy of a charged drop. The last term is the "pairing" energy related to the quantum corrections and the shell effects in the structure of a nucleus. By the order of magnitude, the "pairing" energy is equal to the energy of separation of a neutron from a nucleus $S_n = B(A, Z, D_f) - B(A-1, Z, D_f)$. The coefficients in (1) are usually chosen from the condition of the best fitting of experimental data.

For the first time, the representations about a nucleus as a system revealing the collective behavior and properties arose in connection with the attempt to describe the processes of fission of nuclei. Moreover, the term "fission" appeared in work [5] due to the analogy with the biological process of fission of cells. The description of properties of a nucleus involved the Frenkel model ideas of a charged liquid drop [6] which were developed in the theory of a liquid nuclear drop by Bohr and Wheeler [7, 8] and well agreed with the Bethe – Weizsäcker theory.

Formula (1) was further modified, and its coefficients were corrected on the basis of permanently renewed experimental data [9]. The development of the ideas of nuclei and the nuclear matter within the theory of Fermi-fluid allowed one to calculate the coefficients in the Bethe - Weizsäcker formula and, proceeding from the general representations about the structure of a nucleus, to theoretically determine the binding energy of nuclei with a sufficient accuracy [10, 11].

The phenomenological theory of the binding energy of nuclei on the basis of the drop model leads to a nonmonotonous of the specific binding energy per nucleon on the ratio of the numbers of protons and neutrons (see (1)) and to the existence of the maximum of the specific binding energy per nucleon in the region of nuclei with mass numbers close to those of the stable isotopes of iron and nickel.

As a result, it is traditionally considered that only two types of nuclear processes (reactions) with a positive energy yield (i.e., processes causing the growth of the binding energy of a system):
- reactions of fusion, at which nuclei lighter than iron form heavier nuclei;
- reactions of fission of nuclei heavier than iron into lighter ones.

In any case, according to these ideas accepted also at the present time, the elements in a neighborhood of the *local* maximum of the specific binding energy (elements of the "iron" peak) cannot be used as an efficient source of energy.

However, it becomes more and more clearly now that the ideas of a spatial structure of the dense matter in nuclei should be reconsidered, and the analysis of the variety of possible nuclear structures is required again.

The steady ideas of structures of the nuclear matter do not already correspond to the level of our knowledge and experimental results [12]: "What we have learned over the last decade of research on exotic nuclei forces us to revise some of our basic truths. These were deduced from intensive studies of stable nuclei, but it has become clear that stable isotopes do not exhibit all features… *Nuclear radii don't go as* $A^{1/3}$. For all stable isotopes the density in the atomic nucleus as well as the diffuseness of the surface are nearly constant. Explorations into the far-unstable regions of the nuclear chart have convincingly shown that the diffuseness, and thus the radii of the atomic nuclei, vary strongly… *Many more bound nuclei exist than anticipated.* The neutron drip line is much further out than anticipated twenty years ago. The importance of nucleon correlations and clustering that create more binding for the nuclear system has been underestimated."

The indicated circumstances make it necessary to construct the new more general relations for the calculation of the binding energy of developed nuclear structures. All previous studies and attempts to generalize the Weizsäcker formula (see, e.g., [9, 10]) were based on the application of the theory of analytic functions and the geometry of regular formations. Here, we first make attempt to estimate the coefficients of the Weizsäcker formula with the use of general notions of the fractal geometry. We will demonstrate that the nuclei with more complicated structure than that a drop of the nuclear fluid (in the general case, the nuclei representing fractal nuclear structures) have properties qualitatively different from those of ordinary nuclei. The binding energy of such fractal nuclei increases significantly, and the dependence of the binding energy on the nucleus mass changes qualitatively. In particular, there appears the possibility for the existence of stable superheavy nuclei with specific binding energy exceeding the relevant values characteristic of nuclei of the "iron" peak from the Periodic Mendeleev table, which open new perspectives in the development of the nuclear power industry.

*Geometrical properties of fractal structures*

Among the commonly accepted postulates, the assertions concerning the structure of nuclei and their stability give rise now to the strongest doubts. According to the hydrodynamical model (where a nucleus is represented by a homogeneous spherical drop of the nuclear Fermi-fluid [8] with the mean density of nucleons $\rho \approx 0.17$ nucleon/(fm)$^3$), the number of nucleons in a nucleus, $A$, and the external radius of a nucleus, $R_A$, are connected with each other by the relations

$$R_A(A,\rho) = r_0 A^{1/3}, \quad r_0 = \left(\frac{3}{4\pi\rho}\right)^{1/3}. \tag{2}$$

However, as was indicated above, the contemporary studies of the structure of nuclei show that such ideas of the structure of a nucleus must be reconsidered and reanalyzed from the viewpoint of the extension of the set of elements of the "nuclear Lego constructor." For example, it was assumed in [13] that the superheavy nuclei are clusters of $\alpha$-particles, the possibility of the existence of bubble and quasibubble nuclei was comprehensively studied in [14], and the results of numerical studies of the stability of nuclear structures at low densities indicate the possibility of the existence of nuclear structures [15] which do not correspond obviously to the idea of a liquid drop (see Fig. 1 in [15]).

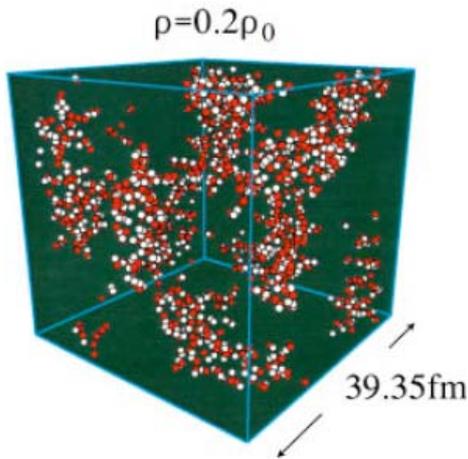 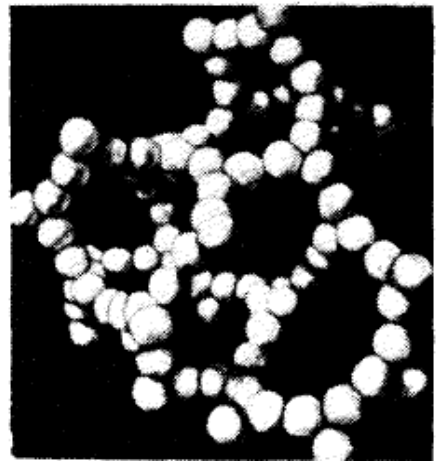

*Fig. 1. Nuclear clusters in the form of a nuclear pasta (nuclear gel) [15]*   *Fig. 2. View of a typical cluster formed at the growth of a structure of solid particles [16]*

In Fig. 2, we present a typical three-dimensional fractal cluster grown due to the Brownian process in the system "particles – cluster" with the probability for particles to stick to the cluster equal to 1 (see [16]). It is seen that the nuclear structure ensuring the minimum energy in the Hartree – Fock approximation constructed in [15] is very close by its spatial distribution to the classical fractal cluster of particles.

The notion of a fractal object was introduced by Mandelbrot [17] into science as an alternative to the regular geometrical and physical objects. In the Nature, we observe a lot of periodic regular phenomena: from the motion of a pendulum to the oscillations of atoms. Despite the absence of strictly periodic motions in the Nature (at least due to the boundedness in time), the periodicity turns out to be the exceptionally useful notion for the explanation of the basic laws and mechanisms in many branches of natural science. One of the reasons for the universality of harmonic motion is the quasilinearity of many physical systems and the invariance of laws guiding their behavior at shifts in space and in time. However, in the course of time, the dominance of linear ideas in science is broken now by the spreading of new nonlinear approaches to the real phenomena surrounding us. The new century has started under the sign of a total penetration of nonlinear phenomena, "nonlinear" thinking, and methods of nonlinear physics into all fields of knowledge.

In the majority of real phenomena, the linearity is violated, and, instead of the periodicity, we deal with aperiodic chaotic motions. In this case, the arising geometrical structures turn out irregular and rugged. At the huge variety of the behavior of nonlinear systems which appear as a result of the nonlinear evolution, there exist the general properties common for most of these systems such as the self-similarity and the invariance relative to a change of the scale (scaling). In other words, the main feature of nonlinear systems is not the invariance under additive shifts, but the invariance under multiplicative transfor-mations of the scale and, hence, the specific role of fractal functions and distributions, rather than that of harmonic ones. Scaling is revealed in many nonlinear physical processes, especially at the study of critical phenomena characteristic of the behavior of substances in a neighborhood of phase transition points.

In the general case, one of the most considerable consequences of the self-similarity is the existence of objects with exceptionally irregular structure, which are called fractals [17]. In some meaning, the self-similarity is also a periodicity, but only on the logarithmic scale. The self-similarity - strict or approximate – plays the principal role in many fields, though it is revealed in very different ways. One of the last branches of physics, where the notion of a fractal is not yet used, is nuclear physics.

In the present work, we will construct a phenomenological theory of the binding energy of nuclear fractal structures, by considering them as the structures determined by their basic geometrical characteristic, namely by the fractal dimension $D_f$.

One of the power dependences characteristic of a fractal is that of the correlation function averaged over angles on the distance to its geometrical center. In this case, the mean density of particles in the cluster $\rho(r)$ varies, as a function of the distance $r$, inside the cluster by the law coinciding with the law of decay of spatial correlations:

$$\rho(r) = \rho_{str}\left(R_{str}(A_{str})/r\right)^{3-D_f}. \tag{3}$$

These general consequences of the fractal geometry yield a simple relation between the mass number of a fractal cluster $A$, external size of the cluster $R_A$, and characteristic size $R_{str}(A_{str})$ of structureless elements, i.e., monomers with the mass number $A_{str}$ and the density $\rho_{str}$, from which the fractal structure is built [16],

$$A = kA_{str}\left(R(A, A_{str}, \rho)/R_{str}(A_{str})\right)^{D_f}, \tag{4}$$

where the coefficient $k$ has value of the order of 1 and is determined by a packing of monomers in the cluster. Relation (3) can be used to present the external radius of the cluster through its mass number and the fractal dimension:

$$R_A(A, A_{str}, D_f) = R_{str}(A/A_{str})^{1/D_f}. \tag{5}$$

As monomers, we can consider any nuclear structures with a sufficiently high stability or/and particles efficiently forming the condensate (particles which can participate in the evolutionary sequence of nuclear phase transitions). We may assume that monomers are, first of all, $\alpha$-particles from the side of low-weight structures (see, e.g., [13]) and the nearest neighbors of helium nuclei in the Periodic Mendeleev table (i.e., stable light nuclei, which can form the condensate of particles with a high probability, e.g., lithium nuclei). From the side of heavier nuclei, the role of monomers can be played by the most stable nuclei such as nuclei of carbon, oxygen, and iron. Moreover, as will be shown below, the most probable monomer for a low-density structure of nucleons with $A \gg 5 \cdot 10^4$ immersed in the electron fluid is iron. As for the fractal dimension, the modeling of the processes of growth of clusters in the three-dimensional space at a high probability of the adhesion of monomers to one another shows that the fractal dimension of clusters is near $D_f = 2.39$.

For the further applications, it is convenient to use the following estimate of the surface area of a fractal cluster depending on the mass number of a structural unit $A_{str}$ and the fractal dimension:

$$S(A; A_{str}, D_f) = 4\pi R_A^2 (A/A_{str})^{\gamma-2/3}. \tag{6}$$

The correction factor indicating the degree of growth of the surface area determined by the internal structure of a cluster depends on the number of structural elements and is equal to $\left(\dfrac{A}{A_{str}}\right)^{\gamma-2/3}$, $\dfrac{2}{3} \leq \gamma \leq 1$. The coefficient $\gamma = 2/3$ for a continuous structure ($D_f = 3$) and $\gamma = 1$ for a developed system of nuclear "threads of a web" or "bubble" nuclei ($D_f = 2$).

Relation (3) yields also the dependence of the fractal dimension of a cluster with the mass number $A$ on the mean density and the mass number of structureless units forming the cluster:

$$D_f = 3\frac{\ln(A/A_{str})}{\ln(A/A_{str}) + \ln(\rho_{str}/\rho)}. \tag{7}$$

The natural attempt to use the conceptions of the fractal geometry in the model of a structure of multinucleon systems must obviously involve their following specific features:

- the appearance of the developed surface (and, hence, a decrease of the binding energy of a nuclear system due to the increase of its surface energy);
- The increase of characteristic sizes of a nuclear conglomerate, as the mean density of the nuclear matter decreases (i.e., a decrease of the Fermi energy of a system and, as it will be shown below, an increase of the binding energy of a nuclear cluster).

We may expect that the formation of the internal fractal structure in nuclear systems leads to a variation of their binding energy in very wide limits.

In the general case, the growth of structures and, hence, a decrease of the number of degrees of freedom of a system, are described by the appearance of two following components of the system: the structureless part and the coherent part in the form of a nuclear "web" constructed from monomers with the mass number $A_{str}$ and some share of protons in monomers $y_{str} = Z_{str} / A_{str}$. In this case, we denote the mass of the coherent part by $m_{cog}$ and the mass of the substance in the structureless liquid part by $m_g$ and introduce the coherence parameter $\eta \approx m_{cog} / (m_{cog} + m_g)$.

The mean density of a system possessing the coherent part varies by the power law (6) due to correlations. Since the potential energy of the nuclear substance is proportional, in general, to the density, we may write

$$U(\rho) \propto U(r), \ U(\alpha r) \propto r^{k_{sc}}, \ k_{sc} = D_f - 3. \tag{8}$$

The value of coherence parameter can be estimated from the Lagrange theorem (virial theorem) for systems with the potential energy possessing the property of similarity, if we take into account that the coherent part is characterized mainly by the potential energy, whereas the whole kinetic energy is present in the structureless (liquid) part. Then the relation of the mean values of kinetic and potential energies

$$\frac{\overline{W_{kin}}}{\overline{U}} = \frac{D_f - 3}{2} \tag{9}$$

yields the formula for the coherence parameter

$$\eta \approx \frac{3 - D_f}{D_f - 1}, \ D_f = \frac{3 + \eta}{1 + \eta}, \ 0 \leq \eta \leq 1. \tag{10}$$

The contribution of the coherent part of the system to the binding energy by the relations $A_{cog} = \eta A$ and $Z_{cog} = y_{str} A_{cog}$, where $A_{cog}$ is the mass number, and the mean density is determined by the fractal dimension (or the coherence parameter) according to relation (3).

*Main contributions to the binding energy of nuclear systems*

As is well known, the interaction of elements of a system causes the effective decrease of the mass of the system (the mass defect appears) with the appearance of the binding energy of the system determined by this mass defect. For example, the total internal energy of a nucleon system including $A$ nucleons ($Z$ protons with mass $m_p$ and $(A-Z)$ neutrons with mass $m_n$) can be written in the form: $W = Z m_p c^2 + (A-Z) m_n c^2 - B$, where $B$ is the binding energy of the system, to which all basic interactions make contributions: $B = B_{Strong} + B_Q + B_{surf}$.

Here,

- $B_{strong} = B_{bulk} + B_{Fermi}$ is the contribution of the strong interaction consisting of two terms: $B_{bulk}$ bearing the bulk character (i.e., $B_{bulk}$ is a strong-interaction-induced part of the binding energy which is proportional to the mass number $A$) and the term $B_{Fermi} \approx -A \frac{3}{5} E_f(\rho)$ which arises due to the Pauli principle and is proportional to $(N-Z)^2$;
- $B_Q$ is the contribution of the Coulomb interaction;
- $B_{surf}$ is the contribution of the surface energy (i.e., energy related to the degree of inhomogeneity of the distribution of nucleons in space).

In order to estimate the contributions to the binding energy, we need the assumptions about the basic geometrical characteristics of a distribution of the nuclear matter in the system. The first efficient model of a distribution of the substance of a nucleus was the simplest model of a drop of the nuclear fluid, in which the mass number $A$ and the radius of a nucleus are connected through relation (2).

At small excitations of a nucleus, the distribution of nucleons in the momentum space is usually considered homogeneous inside of the Fermi spherical surface with the radius in the momentum space equal to the Fermi momentum $p_f$: $f(p, E_f) = \begin{cases} 1, p < p_f \\ 0, p > p_f \end{cases}$ (degenerate Fermi distribution function), as well as in the volume of a nucleus $V$. The integration of the Fermi distributions over the phase space gives the relations between $p_f$ and the densities of nucleons:

$$p_{fn} = (3\pi^2)^{1/3} \hbar \rho_n^{1/3}, \quad \rho_n = \frac{A-Z}{A}\rho, \quad p_{fp} = (3\pi^2)^{1/3} \hbar \rho_p^{1/3}, \quad \rho_p = \frac{Z}{A}\rho. \quad (11)$$

The Fermi surface radius (Fermi momentum) is connected with the Fermi energy $E_f$:

$$E_{f_p} = \frac{p_{fp}^2}{2m_p} = \frac{(3\pi^2)^{2/3} \hbar^2}{2m_p}\rho_p^{2/3}, \quad E_{f_n} = \frac{p_{fn}^2}{2m_n} = \frac{(3\pi^2)^{2/3} \hbar^2}{2m_n}\rho_n^{2/3}. \quad (12)$$

The kinetic energy $W_{kin}(A,Z)$ of the ensemble of nucleons in a nucleus with volume $V$ is determined by their Fermi energy:

$$W_{kin}(A,Z) = 2\frac{V}{(2\pi\hbar)^3}\int\left(\frac{p^2}{2M_{nuc}}\right)\left(f_p(p, E_{f_p}) + f_n(p, E_{f_n})\right)d^3p = \frac{V}{5\pi^2\hbar^3}\left(p_{fn}^3 E_{fn} + p_{fp}^3 E_{fp}\right) =$$

$$= \frac{3}{5}\left((A-Z)E_{fn} + Z E_{fp}\right) = \frac{3}{5}\frac{(3\pi^2\hbar^3)^{2/3}}{2m_n}\rho_{nuc}^{2/3}\left(\left(1-\frac{Z}{A}\right)^{5/3} + \frac{m_n}{m_p}\left(\frac{Z}{A}\right)^{5/3}\right)A. \quad (13)$$

The expansion of $W_{kin}(A,Z)$ in the small parameter $\varepsilon = \left(\frac{1}{2} - \frac{Z}{A}\right)$ leads to the expression

$$W_{kin}(A,Z,\rho_{nuc}) \approx \frac{(3\pi^2\hbar^3)^{2/3}}{2m_n}\rho_{nuc}^{2/3}\frac{2^{1/3}}{4}\left(1+\frac{m_n}{m_p}\right)\left(\frac{3}{5} + \frac{1}{3}\left(1-\frac{2Z}{A}\right)^2\right)A = W_{kinBulk} + W_{kinSym}, \text{ where}$$

$$W_{kinBulk} = \frac{3}{5}\frac{(3\pi^2\hbar^3)^{2/3}}{2m_n}\rho_{nuc}^{2/3}\frac{2^{1/3}}{4}\left(1+\frac{m_n}{m_p}\right)A, \quad W_{kinSym} = \frac{1}{3}\frac{(3\pi^2\hbar^3)^{2/3}}{2m_n}\rho_{nuc}^{2/3}\frac{2^{1/3}}{4}\left(1+\frac{m_n}{m_p}\right)\left(1-\frac{2Z}{A}\right)^2 A. \quad (14)$$

Thus, the formula for the kinetic energy contains the bulk term $W_{kinBulk}$ proportional to $A$ and the term $W_{kinSym}$, which is proportional to $\left(1-\frac{2Z}{A}\right)^2$ and determines the coefficient $c_3$ in the Bethe - Weizsäcker formula (1).

It is clear now that the coefficient $c_0$ in (1) consists of two terms:

$$c_0 \approx U_{eff} - k_{vir}W_{kinBulk} \approx U_{eff} - \frac{3}{5}k_{vir}\frac{(3\pi^2\hbar^3)^{2/3}}{2m_n}\frac{2^{1/3}}{4}\left(1+\frac{m_n}{m_p}\right)\rho_{nuc}^{2/3}. \quad (15)$$

Here, $U_{eff}$ is the depth of the potential well of the strong interaction of nucleons in a nucleus, $k_{vir}$ is the virial coefficient determining the share of the kinetic energy that contributes to the binding energy. The value of $U_{eff}$ can be calculated in the Fermi-fluid approximation, $U_{eff} \approx 58$ MeV (see, e.g., [10]). In this case, the coefficient $c_3$ reads

$$c_3 = \frac{1}{3}k_{vir}\frac{(3\pi^2\hbar^3)^{2/3}}{2m_n}\frac{2^{1/3}}{4}\left(1+\frac{m_n}{m_p}\right)\rho_{nuc}^{2/3}. \quad (16)$$

It is seen that the negative contributions to the binding energy decrease, as the mean density of the nu-clear substance $\approx \rho_{nuc}^{2/3}$ decreases. Therefore, the appearance of fractal structures in the nuclear matter (related to the enhancement of correlations) decreases its mean density and, hence, increases the binding energy of a nuclear structure. We choose $k_{vir}$ so that the binding energy of the known "drop-like" nuclei is maximally exactly approximated. With regard for the geometrical properties of fractals (see relation (3) for the density of nuclear clusters), the above-presented relations yield finally the contribution of the strong interaction in the form

$$B_{Strong} \approx \left(c_0 - c_3\left(1 - \frac{2Z}{A}\right)^2\right)A, \; c_0 \approx \left(58.4 - 42.6\left(\frac{A}{A_{str}}\right)^{-2\left(\frac{1}{D_f} - \frac{1}{3}\right)}\right), \; c_3 \approx 23.7\left(\frac{A}{A_{str}}\right)^{-2\left(\frac{1}{D_f} - \frac{1}{3}\right)}. \quad (17)$$

It follows from (17) that, indeed, the appearance of fractal structures in the nuclear matter (related to the enhancement of correlations and to $D_f < 3$) decreases its mean density (in correspondence with (3)) and, hence, increases the bulk contribution of the strong interaction to the binding energy of a nuclear structure. As $D_f \to 3$, the contribution of the strong interaction tends to the bulk contribution in the ordinary Weizsäcker formula (1).

Since the nuclei are finite systems, there exists also the contribution of the strong interaction related to a great inhomogeneity of a distribution of the nuclear matter near the boundary of the system (surface energy of a liquid drop $W_{surf}(A, \rho_{nuc})$) [10]: $W_{surf}(A, \rho_{nuc}) = \int \lambda_N(\rho_{nuc})(\nabla \rho_{nuc})^2 d^3r$.

For a step-like distribution, this contribution has naturally the form proportional to the surface area: $W_{surf}(A) = \sigma S(A) \approx c_1 A^{2/3}$. Here, the experimental value of the coefficient of surface tension $\sigma$ amounts to about 1 MeV/fm². The fractal structure of a nucleus causes, naturally, a change of the surface of a nuclear structure in agreement with formula (6) for the area of a cluster. With the use of the properties of the fractal geometry, we obtain

$$B_{surf}(A; A_{str}, D_f) = -\sigma S(A; A_{str}, D_f) \approx -c_1(A; A_{str}, D_f) A^{2/3}, \; c_1(A; A_{str}, D_f) \approx 18.56\left(\frac{A}{A_{str}}\right)^{2\left(\frac{1}{D_f} - \frac{1}{3}\right)}. \quad (18)$$

Relations (18) imply that the fractality of a nuclear structure leads to an increase of the surface area (increase of the negative contribution to the binding energy) and, hence, to a decrease of the total binding energy of a cluster.

It is natural that the Coulomb energy of protons in a nucleus is expressed in terms of the nucleon distribution density and the Coulomb potential $U_Q(r, Z) = \frac{Ze}{r}$. For the drop model, the calculations are trivial:

$$W_Q(A, Z, \rho_{nuc}) = \frac{1}{2}\int U_Q(r - r_1, Z)\frac{Ze}{A}\rho_{nuc}(r_1)d^3r_1 = \frac{3}{5}Z(Z-1)\frac{e^2}{R_A} \approx c_2 \frac{Z^2}{A^{1/3}}.$$

It is obvious that the influence of the fractal geometry on this contribution is manifested in its decrease due to an increase of the size of a nucleus. With regard for the dependence of the radius of a fractal cluster on its fractal dimension, we obtain the relation

$$B_Q(A, Z; A_{str}, D_f) = -W_Q = -\frac{3}{5}\frac{e^2 Z(Z-1)}{R_A(A; Z, D_f)} \approx -c_2 \frac{Z(Z-1)}{A^{1/3}}, \; c_2(A, A_{str}, D_f) \approx 0.71\left(\frac{A}{A_{str}}\right)^{-\left(\frac{1}{D_f} - \frac{1}{3}\right)}. \quad (19)$$

The fractality of the structure causes the increase of its external radius, the decrease of the modulus of the negative contribution of the Coulomb energy to the binding energy of a cluster, and, hence, the increase of the total binding energy.

Finally, we arrive at the modified Weizsäcker formula for the binding energy of a nuclear cluster with the mass number $A$, charge $Z$, and fractal dimension $D_f$, which is built of nuclei with the mass number $A_{str}$, as the ordinary Weizsäcker formula (1), but with the coefficients $c_0, c_1, c_2$, and $c_3$, which are functions given by relations (17)-(19), rather than constants. Thus, a variation of the structure of a nuclear cluster leads to a change of the binding energy and the conditions for the stability of nuclear structures.

*Stability of nuclear systems with different structures*

The energy of a nuclear system depends usually only on the number of protons and neutrons in the system. The stable states of nuclei satisfy the conditions of positivity of the binding energy $B$, In this case, the condition of connectivity of neutrons and protons holds automatically, i.e., the relations

$$\mu_n = \left(\frac{\partial W}{\partial N}\right)_Z < 0, \ \mu_p = \left(\frac{\partial W}{\partial Z}\right)_N < 0, \quad (20)$$

where $\mu_n$ and $\mu_p$ are the corresponding chemical potentials of the nuclear system, are valid. These conditions ensure the existence of a potential "well" by the parameters of the system, so that the most stable states are on the bottom of the potential "well" and are determined by the conditions for the appropriate potentials to be zero (derivatives of the binding energy with respect to the relevant parameters). For nuclear fractal clusters, the fractal dimension turns out to be a thermodynamical parameter. Therefore, the most stable nuclear structures are determined by two drip lines:

$$\partial B(A,Z,D_f)/\partial Z = 0, \ \partial B(A,Z,D_f)/\partial D_f = 0. \quad (21)$$

The first equation yields the analytic formula for the drip line (which corresponds to the equilibrium relative to $\beta$-processes):

$$Z_{st}(A,D_f) = A\left(2 + 0.015 A^{2/3}\left(\frac{A}{A_{str}}\right)^{\left(\frac{1}{D_f}-\frac{1}{3}\right)}\right)^{-1}. \quad (22)$$

To find the analytic solution of the second equation in (21) is a very difficult problem. Therefore, we find its solution numerically on the drip line (22) and obtain approximately:

$$D_{fst}(A) = 2.306(\lg(A))^{0.0415}. \quad (23)$$

The drip line (23) indicates the increase of the degree of saturation of stable nuclei by neutrons due to the appearance of spatial structures in a nucleus (and a decrease of the mean density of the nucleus for those structures).

In Fig. 3, we present the dependence of the fractal dimension of a nuclear cluster on its mass number on the drip line, and Fig. 4 shows the dependence of the mean density of the nuclear matter on the mass number on the same line.

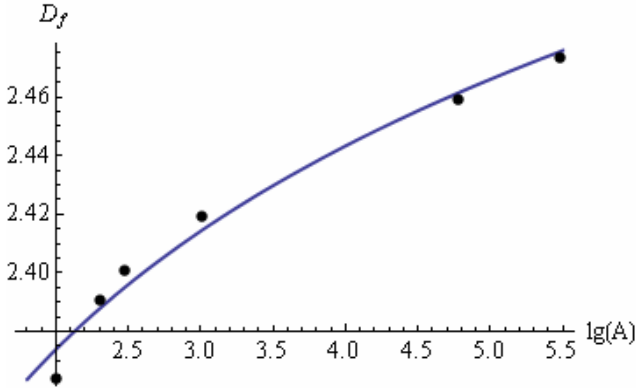 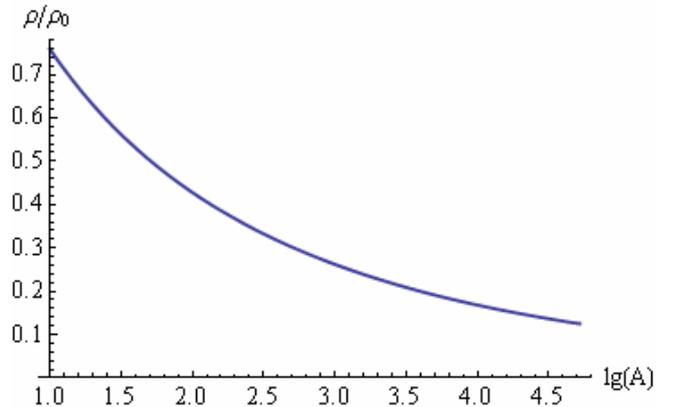

*Fig. 3. Dependence of the fractal dimension on the mass number of a nuclear cluster on the drip line. Points are obtained by the numerical solution of the system of equations (21).*

*Fig. 4. Dependence of the mean density of the nuclear matter on the mass number on the drip line.*

With the use of the drip line (23), relation (22) yields the dependence of $Z/A$ for a stable nuclear structure on its mass number. This dependence is shown in Fig. 5.

Substituting the modified line of $\beta$-stability (22) and relation (23) in (1) with regard for (17)-(19) for the binding energy, we obtain the dependence of the specific binding energy on the mass number. This dependence is shown in Fig. 6. It is seen that the binding energy of stable fractal structures is, firstly, always higher than that of stable structures in the form of liquid drops, and, secondly, the region of stability of nuclear structures becomes wider.

It is necessary to consider the stability of nuclear systems not only relative to the balance of neutrons and protons in a nucleus, but relative to the processes of fission of nuclei. The process of spontaneous fission of a nucleus is hampered by the presence of a potential barrier in the space of parameters which characterize a deformation of a nucleus; the fragments must pass through this barrier prior to their full separation [8].

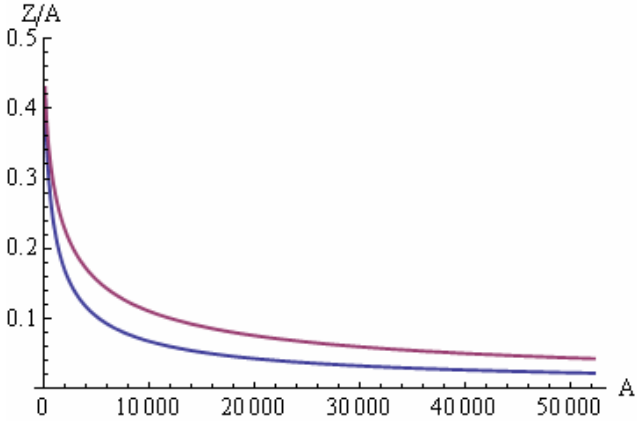
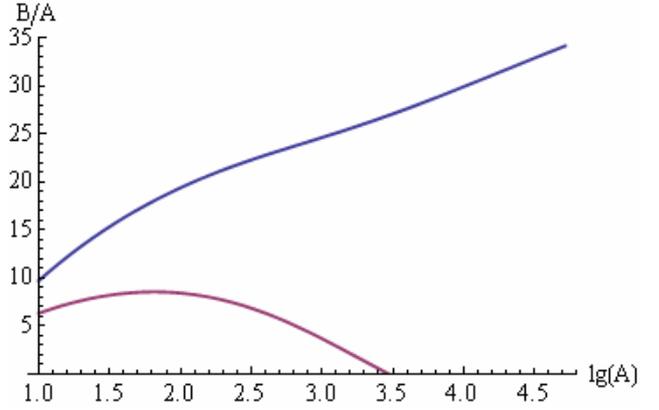

*Fig. 5. Dependence of the relative charge of a nucleus on its mass number (upper curve corresponds to a nucleus in the form of a liquid drop; lower curve – to a fractal cluster)*

*Fig. 6. Specific binding energy of fractal nuclear structures on the drip line. The lower curve corresponds to nuclei in the form of liquid drops.*

In the model of a liquid drop, the fission of a nucleus should be preceded by some deformation. Firstly, an increase of the deformation is accompanied by an increase of the energy of a nucleus. At a deformation of the surface (proportional to a small parameter $\varepsilon$), the Coulomb energy tends to further increase the deformation (by pushing apart the perturbed sections), whereas the surface tension of the drop, which is defined as the derivative of the surface energy, tries to return the spherical shape to the drop. In this case, the energy of a deformation $\Delta W \approx (1/5)(2c_1 A^{2/3} - c_2 Z^2 A^{-1/3})\varepsilon^2$. The condition for the absence of decays on the drip line (positivity of the deformation energy) can be presented in the form $\frac{2c_1}{c_2} > \frac{Z^2}{A}$ and gives the following relation for the limiting value of the mass number $A_b$:

$$A_b \left( 2 + 0.015 A_b^{2/3} \left( \frac{A_b}{A_{str}} \right)^{\left( \frac{1}{D_f} - \frac{1}{3} \right)} \right)^{-2} = 50. \tag{24}$$

The stable nuclei can be observed in the region of mass numbers $A < A_b$. The appearance of a structure in the nuclear matter increases sharply its stability. Large nuclear structures with a sufficiently low density become stable also relative to decays, which is well seen from Fig. 7.

If the conditions of positiveness of the binding energy and the energy of excitation of surface oscillations are satisfied simultaneously, it is possible to determine the boundaries of stability by mass numbers at the given charge of a nucleus. The result of calculations is shown in Fig. 8. The value of mass number for a cluster with the optimum structure and with the maximum binding energy lies between these curves.

The boundaries of stability by the fractal dimension are shown in Fig. 9 as functions of the mass number of a cluster. On the same figure, we show the optimum dimension of such giant fractal clusters.

Work [18] reported on the discovery of stable изотопов of $Th_{90}$ with the mass number 292, which were interpreted as superheavy elements with a charge of 122.

The analysis of Figs. 3, 7, and 8 allows us to propose another interpretation: the observed isotopes are, quite possibly, fractal isomers of nucleus $Th_{90}$, whose large mass number is related to the increase of the number of neutrons in them due to a low mean density of such nuclei, for example, to the formation of spatial structures with the fractal dimension $D_f \approx 2.1$ (quasibubble nuclei).

The possibility of the existence of stable nuclear clusters with the mass composed of tens of thousands of nucleons with a binding energy of about 1 MeV per nucleon and with the sufficiently low stability to the ex-citation of their surface (i.e., with the possibility of a decay induced by external physical actions) allows us to hope that such processes can release significant values of energy. As a result of the experiments on the self-organizing nucleosynthesis in solid targets with the use of hard-current diodes with a special construction [19], a number of processes apparently related to the induced decay were registered in [20,21].

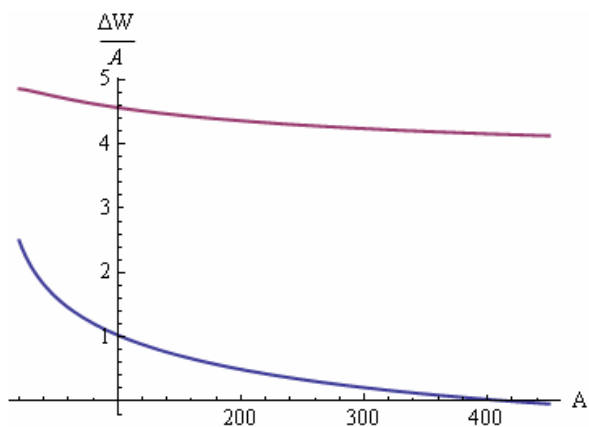 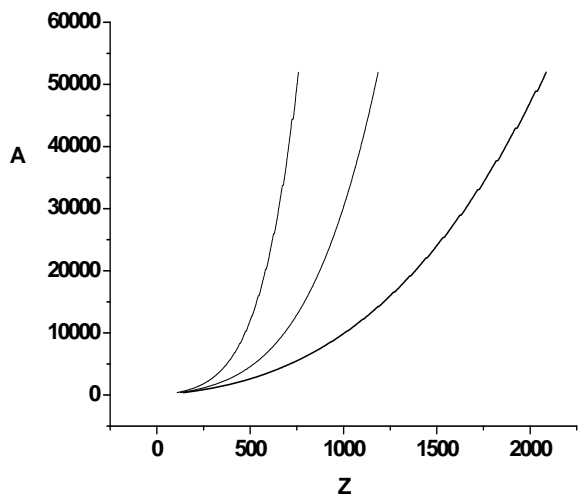

*Fig. 7. Dependence of the energy of a deformation of a nucleus on its mass number. Lower curve corresponds to a liquid drop, and the upper one – to a fractal cluster.*

*Fig. 8. Limiting curves indicating the dependences of the minimum and maximum mass numbers of clusters on the charge of a nucleus. Middle curve corresponds to optimum structures with the maximum binding energy.*

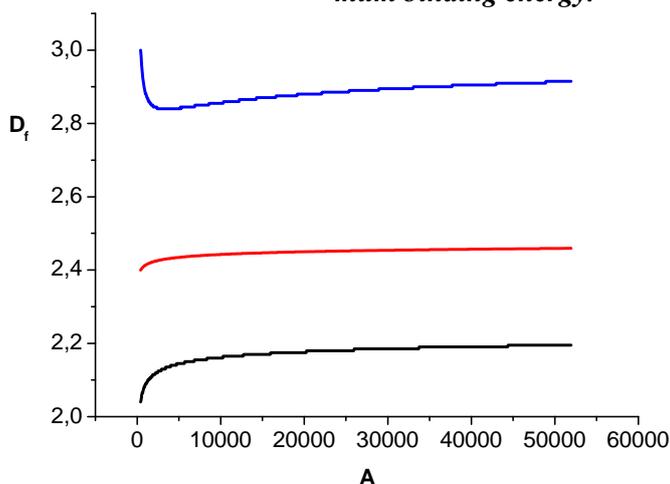

*Fig. 9. Dependence of the limiting values of fractal dimension of a nuclear cluster on the mass number. Middle curve corresponds to the dimension of the cluster with the maximum binding energy.*

For example, with the help of track detectors [22] positioned near the region, where the self-organizing nuclear processes are running, была зарегистрирована the system of tracks of 276 nuclei of lithium and 276 nuclei of helium with an energy of about 1 MeV per nucleon, which escaped from a single center (Fig. 10.).

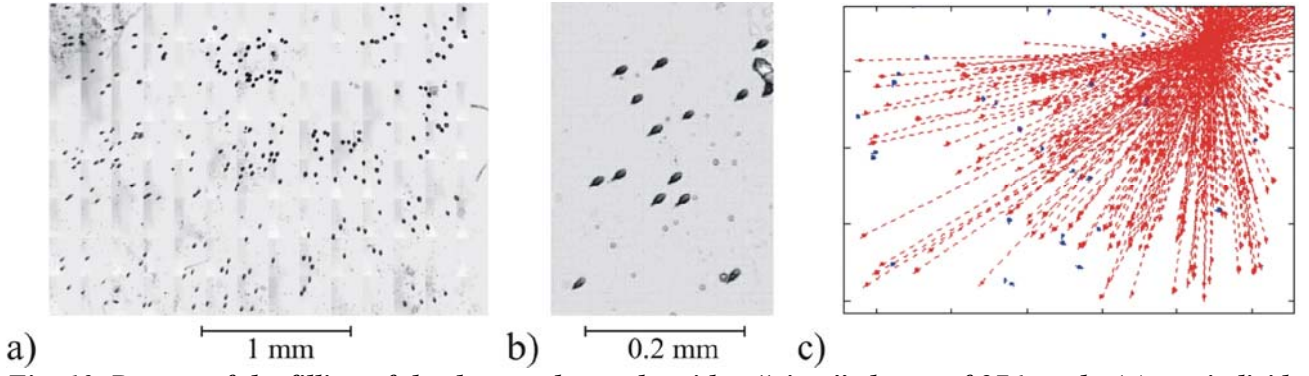

*Fig. 10. Pattern of the filling of the detector by tracks with a "giant" cluster of 276 tracks (a); an individual fragment of the pattern (b); the diagram of track directions (c). Mean energy per particle in the cluster (lithium or $\alpha$-particle) is of the order of 5 MeV.*

The appearance of such tracks can be a result of the spontaneous decay of a giant nuclear cluster of monomers (lithium and $\alpha$- particles) created in the coherent nuclear processes:

$$^{19000}_{1580}A_{D_f=2.69} \rightarrow \,^{16240}_{200}A_{D_f=2.69} + 276\,^{4}_{2}He + 276\,^{6}_{3}Li + W_k \,.$$

We now clarify the above-used designations. The giant nuclear clusters are superheavy nuclear structure, whose composition includes two components: the coherent one in the form of nuclear "threads of a web" with the fractal dimension $D_f$ (its value is given below on the right) formed by monomers with the nuclear density and the noncoherent component. The latter is a structureless part in the form of a nuclear fluid with a relatively low density, where nuclear "threads of a web" are positioned.

The nuclei of lithium and $\alpha$- particles registered in a track detector are monomers, from which the coherent component ($276\,^{4}_{2}He + 276\,^{6}_{3}Li$) is constructed. Together with the nuclear fluid, the nuclear cluster is the structure with the mass number $A_{cl} \approx 19000$, $Z_{cl} \approx 1380$, and $D_f \approx 2.69$, which is denoted by $^{19000}_{1580}A_{D_f=2.69}$. It has a radius of about 45 fm and the stability reserve $\Delta W \leq 0.5$ MeV/nucleon. The kinetic energy released due to the decay, $W_k$, ensures an energy of the order of 1 MeV per nucleon for outgoing fragments. The cluster $^{16240}_{200}A_{D_f=2.69}$ formed after the decay has a higher specific binding energy per nucleon, than the initial structure $^{19000}_{1580}A_{D_f=2.69}$ due to the optimum density of protons in the cluster (see relations (22)-(23)).

The experiments performed much more earlier revealed some anomalies of tracks in nuclear emulsions (see, e.g., [30]), which can be considered as the registration of fractal isomers in the form of quasibubble nuclei with $D_f \approx 2.01$, $A \approx 60$, radius of about 10 fm, and the stability reserve of about 4 MeV/nucleon.

It follows from our studies that there exist the stable nuclei with a high concentration of neutrons and a high binding energy. Therefore, we need to consider the possibilities and the means of creation of stable nuclear structures different from nuclear drops and the set of types of the evolution of multiparticle systems to their equilibrium states.

The solution of this problem is difficult and is far from the completion. However, it is quite obvious that the creation of such superheavy nuclei by the fusion of low-mass nuclei moving with ultra-high energies in direct collisions is a low-efficiency improbable process due to a great excitation of the intermediate nuclear system formed in such collisions.

Works [21, 23-24] proposed a new class of nuclear processes, namely the collective coherent nuclear reactions, which do not require high energies for their realization and occur due to the appearance of long-range correlations in dynamical systems of nuclei with a variable structure. The new synergetic approach to nuclear processes comprehensively presented in [21] is based on such synthetic sciences as the theory of control, self-organization, nonequilibrium thermodynamics of open systems [25-27], and the theory of fractals [17].

The conception of self-organizing synthesis is based on the quite general ideas of a structure of systems and on the comprehension of the fact that the dynamical systems of any nature are not solidified, but "alive" formations revealing the target "behavior." Moreover, their existence is continuously connected with their evolution. According to the conception, the whole observed variety of dynamical systems with various structures is a product of the evolution of the sets of interacting particles on the way of seeking such optimum structure of the system, which would correspond the highest stability and, hence, the largest chances to survive under dominating (i.e., most regular and/or most intense) external actions due to the improvement of the own internal structure.

The conception of self-organizing synthesis, like the scheme of inertial synthesis, considers a nonlinear wave propagating in a medium. In this case, the leading edge of the wave separates naturally the regions of a "fuel" entering into the wave and products of the combustion remaining behind the trailing edge. It is shown in [20, 21] that the analysis of a dynamical system formed by particles of the substance involved in a nonlinear wave leads to the general conclusion about the existence of a possibility to initiate the processes of self-organization in this system, which result in the synthesis of elements due to the "life activity" of the evolving system on the way from its creation to the decay.

By this conception, the realization of any scenario of the synthesis of elements is a collective process or a coherent nuclear reaction, in which a macroscopically large number of nucleons takes participation. A great number of "participants" of such a process corresponds to a huge amount of possibilities to realize the synthesis in the region of stable nuclei (it is proportional to the number of partitions of a very large integer into integer parts) with a minimum number of particles of a "superfluous" building material.

Book [21] contains a comprehensive description of base positions of the conception of self-organizing synthesis of nuclei and the great array of experimental results obtained as a result of the implementation of one of the scenarios of realization of the conception.

One of the most descriptive models of coherent nuclear reactions is the model of the filtration of a flux of initial nuclei through the growing macroscopic nuclear fractal cluster in the form of a moving and evolving shell of the electron - nucleus plasma.

The shell originates, when the correlations attain a critical level at some optimum density, which increases in the course of the evolution of the shell. The growth of the shell density occurs, because a part of the substance of a medium (through which the shell moves) enters the shell structure on its leading edge, whereas a part of the shell evaporates from the trailing edge (relative to its motion; see Fig. 11.).

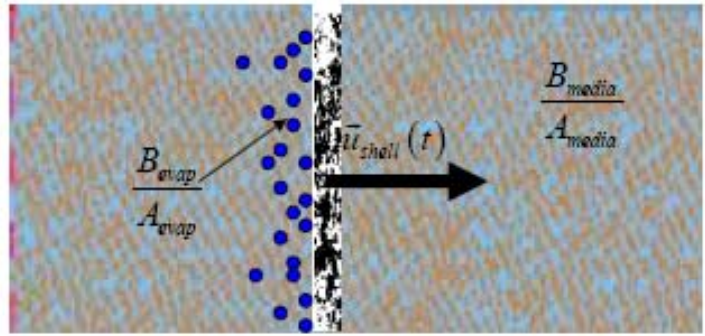

*Fig. 11. Moving fractal shell in the form of a nuclear fractal membrane which filters the medium substance with the formation of new nuclei on the trailing edge of the wave-shell.*

The growth of the shell density is accompanied by a variation of its internal structure, the approaching of the fractal dimension to 3, and a decrease of the stability reserve up to the destruction of the shell and the formation of its fragments with various masses, i.e., up to the synthesis of a spectrum of nuclei.

It is worth noting that the values of fractal dimension of a nuclear cluster (2.46±0.05) presented in Fig. 3 correspond to the typical values of the dimension of clusters growing in the three-dimensional space, which are shown in Fig. 2.

If a cluster is created by the successive attachment of individual particles, its fractal dimension can be determined by the minimization of the free energy of the cluster within a simple model [28]:

$$D_f = \frac{4D_\omega + d(2D_\omega - 4) + 5d^2}{5D_\omega - 4 + 5d}. \qquad (25)$$

Here, $D_\omega$ is the fractal dimension of the trajectories of particles, and $d$ is the space dimensionality. For a Brownian trajectory with the dimension $D_\omega = 2$ in the three-dimensional space, we obtain $D_f = 2.5$, which is close to the dimension of nuclear clusters obtained by us from the condition for the binding energy to be maximum (see Fig. 3).

**Fractal dipole resonance**

The variation of the surfaces of giant charged nuclear clusters (their oscillations) leads, naturally, to the emission. For ordinary nuclei, the emission (giant dipole resonance) can be easily estimated in the approximation of drops of a charged fluid [7, 8]. Let us use this drop model for the estimation of the emission of fractal clusters. We denote a displacement of the surface of a cluster along the radius at a point $(\vartheta,\varphi)$ by $\xi(\vartheta,\varphi) = \xi_0(\vartheta,\varphi)\sin(\omega t)$. The density of the nuclear fluid can be considered constant, $\rho_0$, and let only the form of a nuclear cluster vary. Let the $x$-axis coincide with the polar axis $\vartheta = 0$. The quadrupole moment

$$e\, d_{xy} = \frac{eZ}{V_\delta} \int d\vartheta d\varphi \sin(\vartheta) \int_{R-\delta}^{R+\xi_0(\vartheta,\varphi)} dr\, r^2 r^2 \cos(\vartheta)\sin(\vartheta)\cos(\varphi). \qquad (26)$$

The dependence of a displacement of the surface on the angles $\xi_0(\vartheta,\varphi)$ is described by the expansion in spherical modes (in spherical functions). The lowest nonzero mode is described by the expression $\xi_0(\vartheta,\varphi) = b\cos(\vartheta)\sin(\vartheta)\cos(\varphi)$. Therefore, the quadrupole moment (26) can be written as follows:

$$d_{xy} = \frac{ZR^2}{4\pi\delta} b \int_0^{2\pi} d\varphi \cos^2(\varphi) \int_0^{\pi} d\vartheta \cos^2(\vartheta)\sin^3(\vartheta) = \frac{ZR^2}{15\delta} b \qquad (27)$$

The conservation of the sum of the potential and kinetic energies of surface oscillations implies that the dispersion law of oscillations for the minimum frequency is given by the relation $\omega^2 = 8\left(\dfrac{1}{R}\right)^3 \left(\dfrac{\sigma(A,D_f)}{\rho}\right)$.

In the first approximation (without a consideration of the distributions of protons and neutrons separately), the main contribution to the emission of surface oscillations is made by the quadrupole emission, whose intensity

$$I = e^2 \left(\frac{\omega^6}{4c^5}\right) d_{xy}^2. \qquad (28)$$

The emission line width $\Gamma_\gamma$ is equal to the product of $\hbar$ by the number of emitted quanta:

$$\Gamma_\gamma = \hbar \frac{I}{\hbar\omega} = e^2 \frac{1}{4}\left(\frac{\omega}{c}\right)^5 d_{xy}^2. \qquad (29)$$

Using the dependences of the geometrical parameters of clusters on their dimension and the density, we can obtain the following dependence of the emission frequency of nuclear systems on the mass number, and the fractal dimension:

$$\omega = k^{3/2} \left( \frac{\sigma_s(\rho_A,\delta)}{\rho_A} \left(\frac{A}{A_{str}}\right)^{\gamma-2/3} \right)^{1/2} \approx \frac{\sqrt{8}}{R^{3/2}} \sqrt{\frac{\sigma_s(\rho_A,\delta)}{\rho_A} \left(\frac{A}{A_{str}}\right)^{\gamma-2/3}}. \qquad (30)$$

In Fig. 12, we present the dependences of the emission frequency of nuclei due to the excitation of their surface on the mass number for ordinary nuclei and nuclei with the fractal structure.

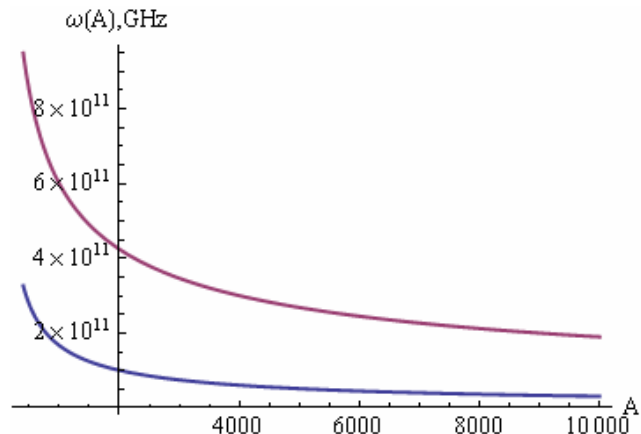

*Fig. 12. Dependences of the emission frequency of nuclei due to the excitation of their surface on the mass number for ordinary nuclei (upper curve) and nuclei with the fractal structure (lower curve).*

Since fractal structures are larger than liquid drops with the same mass, the emission frequencies of the former are significantly less, and the dependences of the emission frequency on the mass number are different. The experimental observation of the dipole resonance of fractal structures will allow one to determine the fractal dimension of nuclei and to classify structural nuclear isomers of the fractal type.

**Fractal Thomson atoms**

In the Nature, the charged nuclei capture the corresponding number of electrons and become the electrically neutral atoms. It is clear that the formed atoms can be stable (in the assumption of their planetary structure), if the radii of the electron orbits closest to the nucleus are larger than the nucleus radius $R(A_M; A_{str}, D_f)$:

$$\frac{a_0}{Z(A_M, A_{str}, D_f)} \geq R_{str}(A_M/A_{str})^{-D_f} \quad (31)$$

This relation between the radii determines, in fact, the boundary of the Periodic Mendeleev table for the mass numbers of elements $A_M$. Relation (31) yields easily the estimate of the parameters of the limiting fractal nucleus: $A_M \approx 52000$, the charge $Z_M \approx 1180$, fractal dimension $D_f = 2.46$, specific binding energy per nucleon is of about 30 MeV, and size $R \approx 90$ fm.

For nuclear systems with large mass numbers ($A > A_M$), the quasistationary electroneutral formations will have the form of stable structures of the electron – nucleus plasma, rather than the planetary atoms. In other words, we may say that they will have the form of fractal Thomson atoms.

For the large values of mass numbers, $A > A_{\max M}$, the fractal electroneutral formations of the electron - nucleus plasma contain obligatorily the contribution of electrons. Their contribution to the energy of the degenerate electron Fermi-fluid can be presented with regard for the quasineutrality as a function of the density nuclear matter as follows:

$$W_{el}(A, Z; D_f) = \frac{3}{4} a_\varepsilon x_\rho^{4/3}(A; D_f)\left(\frac{Z}{A}\right)^{4/3}. \quad (32)$$

The binding energy of such formations differs from the binding energy of nuclei by the contribution of the electron Fermi-fluid. The density of the energy of such formations takes the form

$$\frac{W_{nuclei}(A, Z, x_\rho)}{\rho_{str}} = x_\rho\left((M_n + c_3 - c_0) - (M_n - M_p + 4c_3)\frac{Z}{A} + c_1\frac{1}{A^{1/3}} + c_2\frac{Z^2}{A^{4/3}} + 4c_3\frac{Z^2}{A^2}\right) + \frac{W_{el}}{\rho_{str}}. \quad (33)$$

The limiting Fermi energy of electrons increases with the substance density. Starting from the density which satisfies the condition $Z\varepsilon_e \geq (Am_n - M(A,Z))c^2$, we obtain the possibility for the generation of stable free neutrons under conditions of thermodynamic equilibrium in the reactions $(A,Z) + Ze \rightarrow An + Z\nu_e$.

The requirement of a minimum of the internal energy $w(\rho, \rho_n, A, Z)$ of the nuclear matter with the mean density $\rho$, density of free neutrons $\rho_n$, and density of free electrons $\rho_e$ under the condition of quasineutrality $\rho_e = \frac{Z}{A}(\rho - \rho_n)$ yields the equations

$$\left(\frac{\partial w}{\partial \rho_n}\right)_{\rho, A, Z} = 0; \quad \left(\frac{\partial w}{\partial Z}\right)_{\rho, \rho_n, A} = 0. \quad (34)$$

With regard for (25) after the differentiation with respect to $Z$ and $A$, the conditions of equilibrium (34) yield the equations

$$y_\beta = \frac{Z}{A} = \left(\frac{c_1}{2c_2}\right)^{1/2}\frac{1}{\sqrt{A}}; \quad x_\rho = \left(\left(M_n - M_p + 4c_3\right)\left(\frac{2c_2}{c_1}A\right)^{1/6} - 2\left(\frac{c_1}{2c_2}\right)^{1/3}\left(4c_3\frac{1}{A^{1/3}} + c_2 A^{1/3}\right)\right)^3 \frac{1}{a_\varepsilon^3}. \quad (35)$$

Thus, every density of nucleons corresponds to a single stable nucleus with $A(\rho)$ and $Z(\rho)$. It follows from the solution of Eqs. (35) that the mass number of a stable nucleus tends to the limit $A = 56$, as the den-sity decreases. In this case, $\frac{Z}{A} \to \frac{26}{56}$. The energy density in the electron - nucleus plasma with regard for neutrons is as follows:

$$w = \frac{\rho - \rho_n}{A}\left((A-Z)M_n + ZM_p - B(A,Z)\right) + \rho_n W_n + \frac{3}{4}a_\varepsilon\left(\frac{Z}{A}(\rho - \rho_n)\right)^{4/3}, \quad W_n \approx \rho_n\left(M_n + \frac{3}{5}\frac{a_\varepsilon^2}{2M_n}\rho_n^{2/3}\right). \quad (36)$$

The conditions of equilibrium take the form

$$\left(\frac{2M_n}{a_e^2}\right)^{3/2}\left((c_3 - c_0) + \frac{1}{2}c_1 A^{-1/3} - \frac{2c_1 c_3}{c_2}\frac{1}{A}\right)^{3/2} = x_n \quad (37)$$

$$\frac{1}{a_\varepsilon^3}\left(\frac{2c_2}{c_1}\right)^{1/2}\left((M_n - M_p + 4c_3)A^{1/6} - \sqrt{2c_1 c_2}A^{1/3} - 8c_3\sqrt{\frac{c_1}{2c_2}}A^{-1/3}\right)^3 + x_n = x_\rho. \quad (38)$$

These equations allow one to calculate $A$ and $x_n$ for the given value of density $x_\rho$. As the density increases, the nucleus can be broken down. The threshold value of the ratio $y_{instab} = \frac{Z}{A}$ can be estimated from the requirement that the binding energy of nuclei be zero (see Fig. 11):

$$y_{instab} = \frac{2c_3 - \sqrt{4c_3^2 - (4c_3 + c_2 A^{2/3})(c_3 - c_0 + c_1 A^{-1/3})}}{4c_3 + c_2 A^{2/3}}. \quad (39)$$

The plot of this function is shown in Fig. 13 together with the plot of $y_{st} = \sqrt{\frac{c_1}{2c_2}}A^{-1/2}$ for the most probable (stable) nucleus. The intersection of these plots determines the mass number of the most massive stable drop-like nucleus. In Fig. 14, we show the dependence of the mass number of the most probable nucleus on the logarithm of the mean density of the nuclear matter.

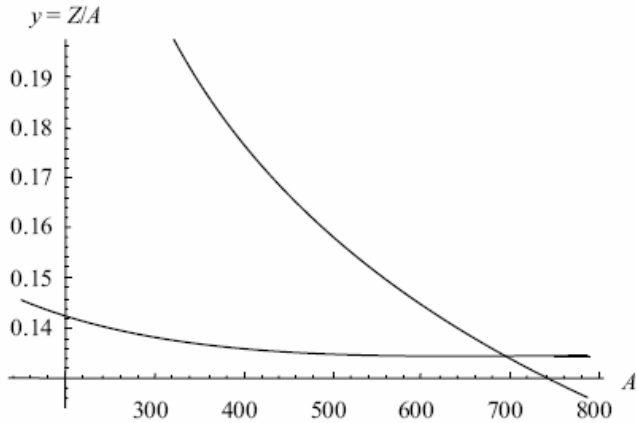
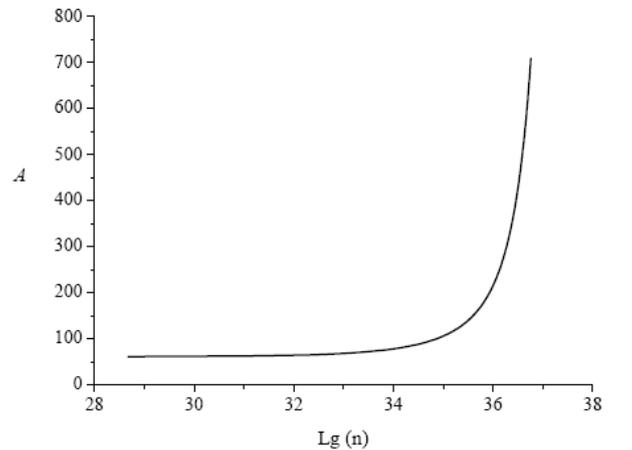

*Fig. 13. Dependence of the quantity $y = Z/A$ on the mass number.*

*Fig. 14. Dependence of the mass number $A$ of the most stable nucleus on the density of nucleons.*

It is seen that the mass number of the most stable nucleus depends weakly on the density of the nuclear matter up to high densities and is equal to 56-60, but then it grows very rapidly.

**Conclusion. Potentialities of the energy release in nuclear processes.**

In the framework of the conception of self-organizing synthesis of nuclei, we have introduced new important notions and presented some results, which would be applied in many fields of science and technique.

We have shown the efficiency of the use of such new states of the nuclear matter as the electron – nucleus plasma with strong correlations and the nuclear gel (cluster condensate with the fractal structure). As a result of the performed studies, it becomes clear that the fractal geometry in the Nature is spread onto the nuclear scale. Together with the mass and charge numbers, the nuclear structures are characterized by such fundamental parameters as the fractal dimension, correlation indices, and critical indices.

Based on the geometrical and physical relations involving the fractal geometry of a cluster structure and the Fermi statistics of nucleons, we have executed the estimates of the binding energy of fractal nuclear structures, predicted a high stability of superheavy nuclear clusters and their high binding energy, and obtained a generalization of the Bethe - Weizsäcker formula for superheavy fractal nuclear structures.

The expansion of the notions of the fractal geometry onto nuclear structures allows us to understand that the potentialities of nuclear processes and technologies as sources of energy are significantly greater than those considered in the modern nuclear physics. The conception of self-organizing synthesis leads us to the global conclusion that the future nuclear technologies will be based on multiparticle collective processes of synthesis-fission (coherent nuclear reactions) with a given energy directedness in a dense coherently correlated plasma (cluster condensate), rather than elementary two-particle collision nuclear reactions with overcoming the Coulomb barrier.

In order to develop the efficient technologies of the release and the accumulation of the nuclear energy, it is necessary to master the control over the self-organization of an internal structure of nuclei and their deformations. The main laws of the self-organization of nuclear structures and the methods of control will be considered in our future publications.